# The Formation and Evolution of Ordinary Chondrite Parent Bodies

**Pierre Vernazza**
*Aix Marseille Université and Laboratoire d'Astrophysique de Marseille*
**Brigitte Zanda**
*Muséum National d'Histoire Naturelle and Observatoire de Paris*
**Tomoki Nakamura**
*Tohoku University*
**Edward Scott**
*University of Hawaii*
**Sara Russell**
*Natural History Museum (London)*

**Abstract**

Ordinary chondrites (OCs) are by far the most abundant meteorites (80% of all falls). Their origin has long been the matter of a heated debate. About thirty years ago (e.g., *Pellas* 1988), it was proposed that OCs should originate from S-type bodies (the most abundant asteroid spectral types in the inner part of the asteroid belt), but the apparent discrepancy between S-type asteroid and OC reflectance spectra generated what was known as the S-type--OC conundrum. This paradox has gradually been resolved over the years. It is now understood that space weathering processes are responsible for the spectral mismatch between S-type bodies and OCs. Furthermore, both telescopic observations and the first asteroid sample return mission (Hayabusa) indicate that most S-type bodies have mineralogies similar to those of OCs. Importantly, the S-type/OC link, which has remained sterile for more than 30 years, has been delivering fundamental constraints on the formation and evolution of planetesimals over the recent years.

## 1. Introduction

Observations of main belt asteroids performed between the early seventies and the late nineties along with meteorite measurements led to the determination of a preliminary version of the compositional distribution in the asteroid belt (see chapter by *DeMeo et al.*; *Gradie and Tedesco* 1982; *Mothe-Diniz et al.* 2003; *DeMeo and Carry* 2013, 2014):

  i) Two main asteroid populations were identfied: the so-called S-types (comprising mostly but not only ordinary chondrite-like asteroids) and C-types (comprising mostly but not only carbonaceous chondrite-like asteroids) accounting for more than 50% of all main belt asteroids and several minor populations (comprising the parent bodies of the remaining meteorite classes and possibly compositions presently unsampled by meteorites);
  ii) A heliocentric gradient was recognized (*Gradie and Tedesco* 1982) with water-poor S-type asteroids being preferentially located in the inner belt while water-rich C-types are the dominant population in the outer belt;
  iii) The existence of a compositional overlap was established (for example, a large number of C-types are currently located where S-types are the most abundant and vice versa).

Since the early 2000s, we have entered a new era of asteroid exploration, mainly because of the emergence of numerous high quality spectroscopic measurements in the near-infrared (obtained essentially with the NASA IRTF ground based telescope) but also because of the Hayabusa asteroid sample return mission (*Nakamura et al.* 2011). Near-infrared spectra for several hundreds of near-Earth (see chapter by *Binzel et al.*; *Thomas et al.* 2014) and main belt asteroids (e.g., *DeMeo et al.* 2009, *Vernazza et al.* 2014) have now been collected. These new datasets allowed us to enter the second phase of the exploration of the compositional distribution of the asteroid belt and of the compositional characterization of the near-Earth asteroid population. While previous measurements in the visible allowed us to establish for example, that ordinary chondrites (OCs) formed closer to the sun than carbonaceous chondrites (CCs), measurements in the near-infrared – which offer significantly more constraints on asteroid compositions than in the visible range alone (see chapter by *Reddy et al.*) – allow us to constrain the respective formation location of the various meteorite classes (H, L, LL, CI, CM, etc..) within these broad groups (OCs, CCs), namely whether for example H chondrites formed closer to the sun than LL ones or vice versa (*Vernazza et al.* 2014). Near-infrared measurements also allow us to attack important questions such as determining the source regions of both meteorites and NEAs (e.g., *Vernazza et al.* 2008, *Thomas and Binzel* 2010).

In parallel, samples of the S-type asteroid Itokawa were brought back to Earth by the Hayabusa mission, which demonstrated that OCs originate from S-type asteroids and validated the numerous laboratory space weathering experiments that had been performed on various meteorite classes (including ordinary chondrites). In particular, these experiments had predicted that space weathering processes were responsible for the observed spectral mismatch between OCs and S-type asteroids. As a matter of fact, this work resolved the S-type – OC conundrum (*Chapman* 2004) that lasted for more than 30 years and demonstrated that S-type asteroids are mostly OCs or OC-like and not differentiated asteroids, as was advocated by *Bell et al.* (1989) and *Gaffey et al.* (1993).

The aim of the present chapter is to summarize the evidence concerning the origin of OCs and discuss the constraints that were accumulated over the years regarding the formation and evolution of their parent bodies (S-type asteroids). While we give appropriate space to the resolution of the S-type--OC conundrum, the main emphasis of the chapter is put on what was learned from the recently established spectroscopic links between specific ordinary chondrite classes (H, L and LL) and their S-type asteroid parent bodies. We will show that these links, alongside new laboratory work on OCs, have not only allowed us to gain a better knowledge of the formation mechanism of ordinary chondrite parent bodies and planetesimals in general, but they also boosted our understanding of the collisional and dynamical evolution of asteroids.

The structure of this review chapter is as follows. We start by reviewing in section 2 the most important constraints on the formation and early evolution of the ordinary chondrite parent bodies and sometimes more generally of the solar system that were derived from laboratory studies of ordinary chondrites. Then, in section 3, we will review the outstanding efforts that have been accomplished over the last 40 years by a large fraction of the asteroid-meteorite community to unambiguously identify the parent bodies of ordinary chondrites. In section 4, we will review the constraints on the formation and evolution of ordinary chondrite parent bodies that were obtained from telescope observations of S-type asteroids and which nicely complement the laboratory constraints presented in section 2. In section 5, we will review the remaining issues in the OC/S-type field that require critical new data.

## 2. Constraints derived from ordinary chondrite studies on the formation and early evolution of their parent bodies

Ordinary chondrite meteorites (OCs) are silicate-rich meteorites composed mostly of olivine and low calcium pyroxene that are by far the most abundant meteorites (80% of all falls; *Hutchison* 2004). They have been subdivided into three groups (H, L and LL) based on variations in bulk composition, such as molecular ratios [FeO ⁄ (FeO+MgO)] in olivine and pyroxene (*Mason* 1963, *Keil & Fredriksson* 1964) and the ratio of metallic Fe to total Fe (*Dodd et al.* 1967) (Table 1). The L group slightly outnumbers the H group among falls (Table 1), but the reverse is true among finds. The LL group is distinctly less abundant than either group (*Hutchison* 2004).

H chondrites have the highest total abundance of Fe (and the highest Fe/Si atomic ratio) among OCs (H stands for « **H**igh » Fe). They are also the most reduced OCs, their iron being mostly in metallic form. When equilibrated, they exhibit the lowest FeO contents in their silicates. At the other extreme, LL chondrites contain the least total Fe, corresponding to the lowest Fe/Si atomic ratio, and they are the most oxidized OCs, with little metal and with the silicates that have the highest FeO contents when equilibrated. They are « **L**ow » in total iron, « **L**ow » in metal, hence the LL denomination. H and LL chondrites also differ in terms of oxygen isotopes, with LL chondrites containing a very small but measurable depletion of $^{16}$O with respect to the other two isotopes compared to H chondrites. L chondrites (**L**ow in total iron) are intermediate between H and LLs in all these properties – somewhat closer to LLs in terms of total Fe.

Table 1 - Properties of equilibrated ordinary chondrites.

|  | **H** | **L** | **LL** |
|---|---|---|---|
| **Fall statistics (%)** | 34 | 37 | 9 |
| **Fe (wt%)** | 28 | 22 | 19 |
| **Fe/Si (atomic)** | 0.81 | 0.57 | 0.52 |
| **Metal (vol%)** | 8.4 | 4.1 | 2 |
| **Fa content of olivine*** | 16-20 | 21-26 | 27-31 |
| **Fs content of pyroxene*** | 15-17 | 18-22 | 22-30 |
| **~Δ$^{17}$O** (‰)** | 0.7 | 1.1 | 1.3 |
| **Ol / (Ol + Px)*** | 51-60 | 60-67 | 70-82 |

* Olivine (Ol) is (Fe,Mg)$_2$SiO$_4$ ; pyroxene (Px) is (Fe,Mg,Ca)$_2$Si$_2$O$_6$. Fayalite (Fa) content of olivine and ferrosilite (Fs) content of pyroxene correspond to the relative abundance of Fe atoms compared to other cations: Fe / (Fe + Mg) and Fe / (Fe + Mg + Ca) respectively. Ol / (Ol + Px) values are after *Vernazza et al.* (2008) and after *Dunn et al.* (2010). The properties listed above are those of equilibrated (metamorphosed) ordinary chondrites. Fa content of olivine and Fs content of pyroxenes are highly variable in unequilibrated chondrites.

**Δ$^{17}$O measures the distance to the terrestrial fractionation line in the oxygen 3-isotope plot. It is defined as Δ$^{17}$O = ∂$^{17}$O-0.52×∂$^{18}$O, where ∂$^{17}$O and ∂$^{18}$O measure respectively the excess of $^{17}$O and $^{18}$O with respect to $^{16}$O and to a standard (Standard Mean Ocean Water), multiplied by 1000.

Textural variations and corresponding mineral and chemical trends indicate that differing degrees of thermal metamorphism (heating) took place within each ordinary chondrite group (*Van Schmus & Wood* 1967; *Dunn et al.* 2010). Based on these variations, a petrologic classification scheme (*Van Schmus & Wood* 1967) for OCs was developed (H, L and LL groups are further subdivided into fours groups – from 3 to 6), which consisted in distinguishing the less metamorphosed (heated) chondrites (type 3, called unequilibrated ordinary chondrites: UOCs) from chondrites that have undergone higher degrees of thermal metamorphism (type 4 to 6, called equilibrated ordinary chondrites: EOCs; see *Huss et al.* 2006 for a review) (Figure 1). Type 3s are further divided into 3.0-3.9 based on induced thermoluminescence properties (*Sears et al.,* 1980), but also by textures of opaque minerals (*Bourot-Denise et al.,* 1997) and the structure of organics as seen in Raman spectroscopy (*Bonal et al.,* 2006; *Bonal et al.,* 2007). An even finer division of petrographic types <3.2 was defined based on the distribution of $Cr_2O_3$ within olivine (*Grossman and Brearley,* 2005).

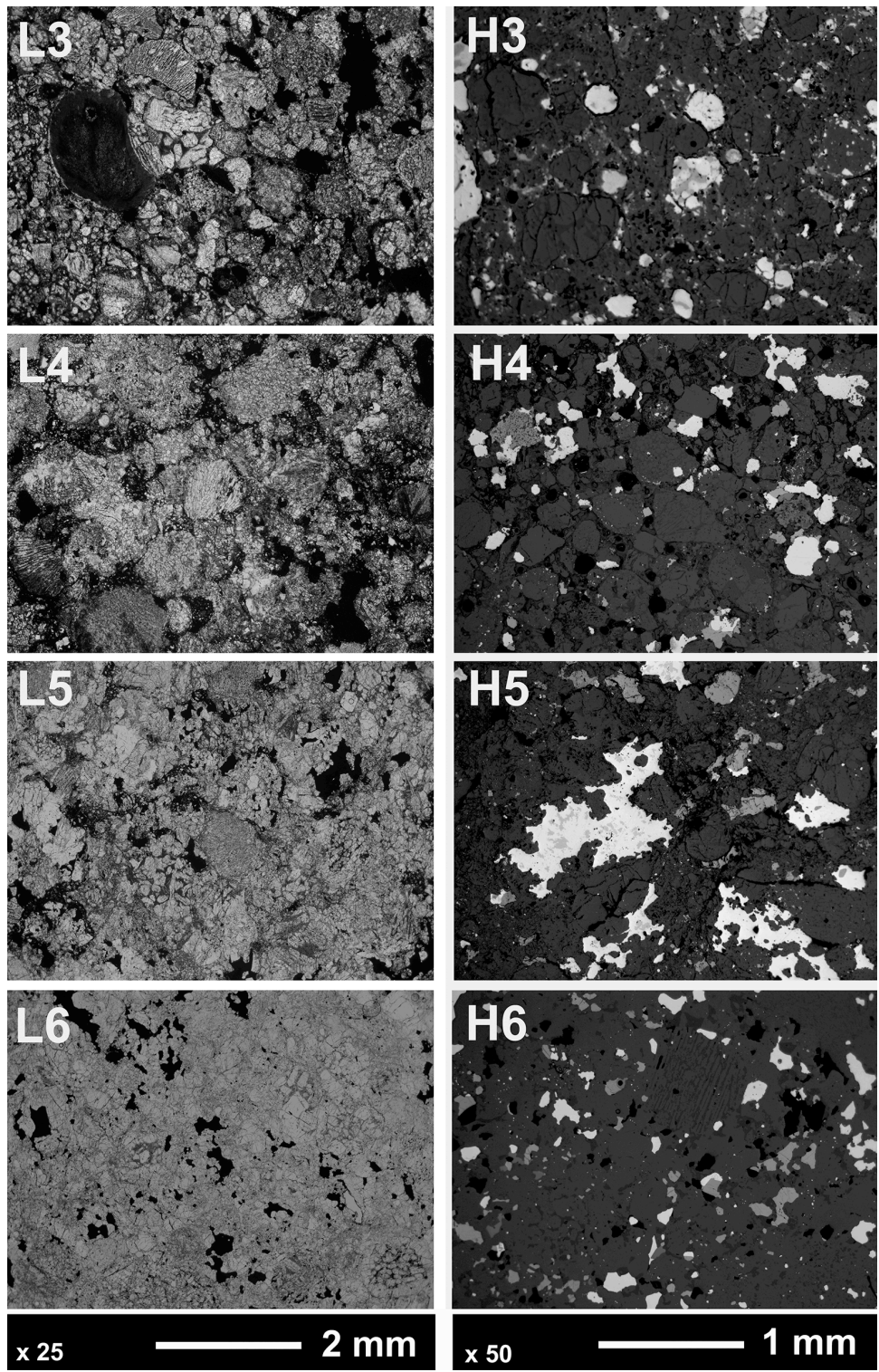

*Fig. 1. Compositional and textural evolution of H and L chondrites as a function of petrologic type* (silicate textures seen in transmitted light in L3-L6 chondrites and opaque mineral textures seen in reflected light in H3-H6 chondrites). *Metamorphic heating in OCs results in chemical*

*and textural changes, the most spectacular of which is the progressive disappearance of chondrules and matrix from petrologic type 3 (least heated) to 6 (most heated). As a result of long duration heating (millions of years), the mineralogy of the rock is modified. Elemental diffusion allows minerals in chemical disequilibrium at the onset to react with one another. Minerals thus "equilibrate", e.g. all olivine grains in one given rock end up with the same (FeO-rich) composition (Table 1). Glass disappears while feldspar grows. Textures also tend to equilibrate with the disappearance of the smaller grains (Ostwald ripening). This makes chondrule outlines become blurry and the typical chondritic texture hence tends to disappearx.*

The study of ordinary chondrites and of other chondrite classes has provided numerous constraints on the formation and early evolution of planetesimals, including 1) the processes that occurred in the protoplanetary disk prior to primary accretion (i.e. planetesimal formation) and their associated timescales (*Cuzzi et al.* 2001, *Cuzzi & Weidenschilling* 2006), 2) the heating mechanisms at the origin of metamorphism in planetesimals (*Ghosh et al.* 2006), and 3) the post- (and syn-) accretional heating and collisional events (*Haack et al.* 1996 and references therein, *Hutchison* 2004, *Huss et al.* 2006, *Ghosh et al.* 2006, *Scott et al.* 2014).

2.1 Constraints on the processes that occurred in the protoplanetary disk prior to accretion

UOCs, which represent ~15 % of all OCs (*Hutchison* 2004), are the most primitive OCs. Several thermometers indicate that a number of UOCs did not experience metamorphic temperatures (*Hutchison* 2004) > ~370 ºC and that all were < ~600 ºC. As such, UOCs, give us the best indication of the initial material (a sort of snapshot of the protoplanetary disk) from which their parent asteroids accreted, as well as key constraints on the events that occurred in the protoplanetary disk prior to primary accretion.

UOCs are mainly aggregates of high-temperature (> 1600 ºC) components, including chondrules (~80% of the volume) which are millimeter-sized silicate spherules that formed through rapid melting and cooling of precursor material, via a still elusive mechanism (*Connolly and Desch* 2004; *Jacquet et al.* 2012), as well as metal and sulfide grains. All these components are set in a fine-grained interchondrule matrix (10-15 vol% of the rock) (*Hutchison* 2004), which is believed to have always remained at low-temperature (< 200 ºC) in the disk.

As of today, the most likely explanation for the simultaneous presence in ordinary chondrites of low-temperature and high-temperature components believed to have formed respectively far from and close to the Sun is that radial mixing was extensive in the solar nebula and thus played a prominent role in shaping the composition of these chondrites and that of planetesimals overall. Indeed, evidence of the simultaneous presence of low-temperature and high-temperature components is also observed in other chondrite classes (*Hutchison* 2004 and references therein) and in comets (*Kelley & Wooden* 2009 and references therein).

Further evidence of the importance of radial mixing in all classes of chondrites, and ordinary chondrites in particular, is provided by i) the 'universal shape' of the size distribution of chondrules that is centered at different sizes from chondrite class to chondrite class (*Grossman et al.* 1989 and references therein, *Cuzzi et al.* 2001, 2008), indicative of size-sorting and ii) the simultaneous presence in a given chondrite of various proportions of two different types of chondrules (reduced type I and oxidized type II chondrules – made in different environments, *Zanda et al.* 2006).

Recently, *Fu et al.* (2014) have provided new constraints on the early evolution of the solar system by measuring the remnant magnetization in olivine-bearing chondrules from the primitive Semarkona meteorite (LL3.0). They report that these chondrules were magnetized in a nebular field of 54 ± 21 µT. This intensity supports chondrule formation by nebular shocks or planetesimal collisions rather than by electric currents, the x-wind, or other mechanisms near the sun. This implies that background magnetic fields in the terrestrial planet-forming region were likely 5-54 µT, which is sufficient to account for measured rates of mass and angular momentum transport in protoplanetary disks.

2.2 Heating mechanisms leading to metamorphism in chondrites

Whereas petrographic types were first discovered and described in OCs, it appears that most chondrites show evidence of metamorphism. Indeed, most enstatite chondrites have petrologic types that range from 4 to 6 with only two unequilibrated objects among observed falls (EH3). Carbonaceous chondrites also show evidence of metamorphism: CK chondrites have petrologic types that range from 3 to 6 and a few CO and CV chondrites may be somewhat metamorphosed, although most of them are ranked type 3s.

Several heat sources have been proposed to explain the thermal evolution (including metamorphism and/or differentiation) of planetesimals (see chapter by *Scott et al.* for a more detailed discussion) and thus the existence of the different petrologic types (3 to 6) within each OC class (see *McSween et al.* 2002, *Ghosh et al.* 2006, *Sahijpal et al.* 2007 for reviews). These include the decay of short-lived radioactive nuclides (*Urey* 1955), which is now considered the most likely heat source, the decay of long-lived radioactive elements (*Yomogida & Matsui* 1984), electromagnetic induction heating (*Sonnett et al.* 1968; *Menzel & Roberge* 2013) and impact heating (e.g., *Rubin* 1995, 2003, 2004). Recent work has shown that neither electromagnetic induction nor impacts alone can explain the thermal processing of planetesimals [see *Marsh et al.* (2006) concerning electromagnetic induction and *Keil* (1997) and *Ciesla et al.* (2013) for impact heating]. There is also no coherent scenario that could favor the decay of long-lived radioactive elements as the only heat source.

More specifically, substantial isotopic evidence suggests that $^{26}$Al was the major source of heat for melting early-formed planetesimals and heating bodies that accreted later (e.g., *Bizzarro et al.,* 2005; *Kleine et al.,* 2005; *Hevey and Sanders,* 2006; *Ghosh et al.,* 2006; *Kruijer et al.,* 2012; *Sanders and Scott,* 2012). However, there is also evidence in chondrites for localized impact heating during metamorphism (*Rubin,* 1995, 2004). Impact heating of asteroids, even porous ones, is very inefficient (*Keil et al.,* 1997), but *Davison et al.* (2012) inferred that large projectiles impacting into highly porous chondritic targets could have buried enough impact melt at depth to cause localized slow cooling. In addition, impacts during metamorphism may have excavated hot chondritic rock so that it cooled more rapidly and mixed hot and cold material (*Davison et al.,* 2013; *Ciesla et al.,* 2013).

Over the last ~30 years, several groups have developed a wide range of thermal models of planetesimals based primarily on $^{26}$Al heating (e.g., *Miyamoto et al.* 1981; *Miyamoto* 1991; *Grimm and McSween* 1993; *Bennett and McSween* 1996; *Akridge et al.* 1998; *Merk et al.* 2002; *Ghosh et al.* 2003; *Tachibana and Huss* 2003; *Trieloff et al.* 2003; *Bizzarro et al.* 2005; *Baker et al.* 2005; *Mostefaoui et al.* 2005; *Hevey and Sanders* 2006; *Sahijpal et al.* 2007; *Harrison & Grimm* 2010; *Elkins-Tanton et al.* 2011; *Henke et al.* 2012a,b; *Neumann et al.* 2012, *Monnereau et al.* 2013). Some authors have included additional heating effects due to $^{60}$Fe, which has a half-life of 2.6 Myr, cf. 0.72 Myr for $^{26}$Al. However, the abundance of $^{60}$Fe is now thought to have been too low to provide any significant heating (*Tang and Dauphas,* 2012).

If the parent bodies of the ordinary chondrites had been heated by $^{26}$Al and cooled without impact disturbance, they would have formed an onion-shell structure in which the most metamorphosed type 6 material occupying the central region surrounded by successive shells of less-metamorphosed type 5 through 3 material (*Trieloff et al.* 2003; *Ghosh et al.* 2006; *Henke et al.* 2012a,b; 2013 and references therein).

2.3 Thermal and impact history of the OC parent bodies

In this section we summarize the evidence that has been accumulated over the years regarding both the thermal and impact history of OC parent bodies.

2.3.1 Evidence for impact from breccias and cosmic-ray exposure ages

About 20-30% of all ordinary chondrites are fragmental breccias composed of various petrologic types (*Bischoff et al.,* 2006). Regolith breccias, which are fragmental breccias containing a small fraction of grains with solar-wind gases that were exposed on the surface of the parent asteroid, account for 15, 3, and 5% of H, L, and LL chondrites, respectively. Fragmental breccias that lack solar wind gases account for 5, 22, and 23% of H, L, and LL chondrites, respectively (*Rubin et al.,* 1983). The fraction of LL chondrites that are breccias may be much higher as *Binns* (1967) found that 62% of LLs were breccias. These breccias show that each OC parent body contains both unequilibrated and equilibrated petrologic type material and that type 3 to 6 material can be found in a single body.

Regolith breccias and gas-poor fragmental breccias are dominated by clasts of equilibrated material. Although a few OC regolith breccias contain predominantly type 3 materials, most are largely composed of equilibrated, type 4-6 material. A typical regolith breccia contains clasts of equilibrated type 4-6 material, with rare impact melt clasts in a dark matrix that is mostly fine equilibrated material with a small amount of type 3 material including some unequilibrated chondrules, (e.g., *Metzler et al.,* 2011). This suggests that the surfaces of the OC parent bodies are dominated by equilibrated material and that each body contains a variety of different types. Few breccias contain more than a tiny fraction of foreign material. This is mostly carbonaceous chondrite material, presumably from projectiles.

Interestingly, the range of composition of olivine grains in a section of matrix from a regolith breccia (e.g., Fa$_{10-29}$ in the NWA 869 L chondrite; *Metzler et al.,* 2011) is comparable to the range observed in the Itokawa sample, Fa$_{24-31}$; *Nakamura et al.,* 2011). In both cases more than 90% of the grains are derived from equilibrated material.

The presence of solar-wind gas and solar-flare tracks in the regolith breccias shows that these breccias were formed after metamorphism of the parent bodies, which would have removed both features. Some gas-poor fragmental breccias may have formed earlier during metamorphism. The L breccia Mezo-Madaras, which contains L4 clasts in a predominantly L3 matrix, probably formed at this time (*Scott et al.,* 2014).

Clues to the recent impact history of the OC parent bodies can be obtained from their cosmic-ray exposure ages, which range from ~1-100 Myr (*Herzog and Caffee,* 2014). Cosmic rays penetrate a meter of rock so these ages date the time when meter-sized meteoroids were formed by impacts on much larger bodies. The spectra of ages for OCs are dominated by a few peaks showing that the production of meteorites is dominated by a very small number of impact events on a correspondingly small number of parent bodies (*Herzog and Caffee,* 2014). H, L, and LL chondrites have different peaks in their age spectra, but petrologic types of the same group show similar peaks, and regolith breccias have similar distributions to other chondrites. These features together with the abundance and nature of breccias suggest that the ordinary chondrite parent bodies were well mixed internally by impacts with all types jumbled up. If the OC parent bodies did once have onion-shell structures they no longer do. This conclusion is consistent with the results of *Vernazza et al.* (2014) discussed below that show metamorphosed material to be present at the surface of H chondrite parent bodies.

Finally, about half of all H chondrites have exposure ages of 6-10 Myr. This peak in the cosmic-ray exposure age spectrum is broader than other OC peaks and it may result from multiple impacts. Another group of meteorites, the acupulcoites and lodranites, have a similar cosmic ray exposure age range of 5-10 Myr (*Terribilini et al.,* 2000 ; *Herzog and Caffee,* 2014). Numerous impacts on different bodies during this period may result from a nearby major impact that produced a family of fragments that peppered nearby asteroids. Thus the 6-10 Myr H chondrite peak should not be taken as evidence that all these meteorites come from one body.

2.3.2 Constraints on the thermal evolution and impact history of the H chondrite parent body from cooling rates

Here we summarize what has been learned about the metamorphic and early impact histories of the H chondrites in relatively unshocked H3–6 chondrites.

*Pellas and Storzer* (1981) found that cooling rates for six H4–6 chondrites derived from Pu fission track thermometry decreased with increasing petrologic type. They concluded that the parent body of the H chondrites had been heated by $^{26}$Al decay and had cooled with an onion-shell structure.

Cooling rates of over 30 ordinary chondrites have also been inferred from Ni concentrations at the centers of zoned taenite grains (*Wood,* 1967; *Willis and Goldstein,* 1981, 1983). *Scott and Rajan* (1981) and *Taylor et al.* (1987) found no evidence that the parent bodies had cooled with an onion-shell structure as cooling rates did not decrease systematically from type 3 to type 6. They concluded that either the parent bodies never had onion-shell structures or their interiors were rearranged by impacts, possibly by catastrophic disruption and reassembly, as *Grimm* (1985) suggested, prior to cooling below ~600 ºC when taenite grains start to develop Ni concentration gradients. Because metallographic cooling rates for the six H chondrites studied by *Pellas and Storzer* (1981) were consistent with fission track cooling rates, *Taylor et al.* (1987) suggested that the apparent conflict between the two techniques was merely a result of the small sample size of the fission track data set.

The most detailed evidence for the onion-shell model was presented by *Trieloff et al.* (2003), who measured Ar–Ar ages and fission track cooling rates for nine H chondrites and compared them with the Pb–Pb ages of *Göpel et al.* (1994). Both sets of radiometric ages were found to correlate inversely with petrologic type and closely matched ages calculated for various depths in a 100 km radius asteroid that was heated by $^{26}$Al to 850 ºC in its central regions and cooled without disturbance by impact to ~100 ºC in ~150 Myr. *Trieloff et al.* (2003) did not discuss the conflicting metallographic cooling rates of *Taylor et al.* (1987) but suggested that the samples studied by these authors had been modified by shock or impact heating. *Wood* (2003) concluded that *Trieloff et al.* (2003) had resolved the controversy in favor of the onion-shell model.

Subsequent studies of the radiometric ages of H chondrites or thermal modeling of their parent body have largely endorsed the conclusions of *Trieloff et al.* (2003). *Kleine et al.* (2008) found that Hf–W ages of five H4–6 chondrites including four of those studied by *Trieloff et al.* (2003) decreased with increasing petrologic type. They concluded that the parent body had cooled with an onion-shell structure after heating by $^{26}$Al. *Harrison and Grimm* (2010) used radiometric ages and metallographic cooling rates for H chondrites to constrain various thermal models and concluded that nearly all data could be accounted for by an onion-shell body of radius 75–130 km. They questioned the accuracy and validity of some metallographic cooling rates and suggested that the relatively small number of non-conforming data might be attributed to impacts that left the parent body largely intact. Finally, *Henke et al.* (2012) and *Monnereau et al.* (2013) focused on the eight H chondrites for which precise radiometric ages are available at three different closure temperatures (*Trieloff et al., 2003*). They concluded that an onion-shell body with a radius 110–130 km that accreted rapidly (in <0.2 Myr) around 1.8–2.1 Myr after CAI formation could successfully reproduce their cooling histories without invoking any disturbance by impacts.

Constraints on the thermal histories of ordinary chondrites from silicates and oxides suggest that the parent bodies did not cool with an onion-shell structure. Cation ordering temperatures in orthopyroxene in eight H4–6 chondrites indicated similar cooling rates at ~400–500 ºC (*Folco et al.* 1996). Olivine-chromite thermometry gave similar mean temperatures for H4, H5 and H6 chondrites of ~690–770 ºC (*Wlotzka,* 2005). *Kessel et al.* (2007) also found several inconsistencies with the onion-shell model. *Ganguly et al.* (2013) measured compositional profiles across coexisting orthopyroxene–clinopyroxene, olivine-spinel, and orthopyroxene-spinel grains in five H4–6 chondrites and found all grains to be essentially homogeneous. From calculated compositional profiles, they inferred that the chondrites cooled very rapidly above 700 ºC at rates of 25–100 ºC/kyr, several orders of magnitude faster than the rates generally inferred from radiometric ages and metallography. *Ganguly et al.* (2013) concluded that the H chondrite parent body had been fragmented and then reaccreted prior to cooling below 700 ºC.

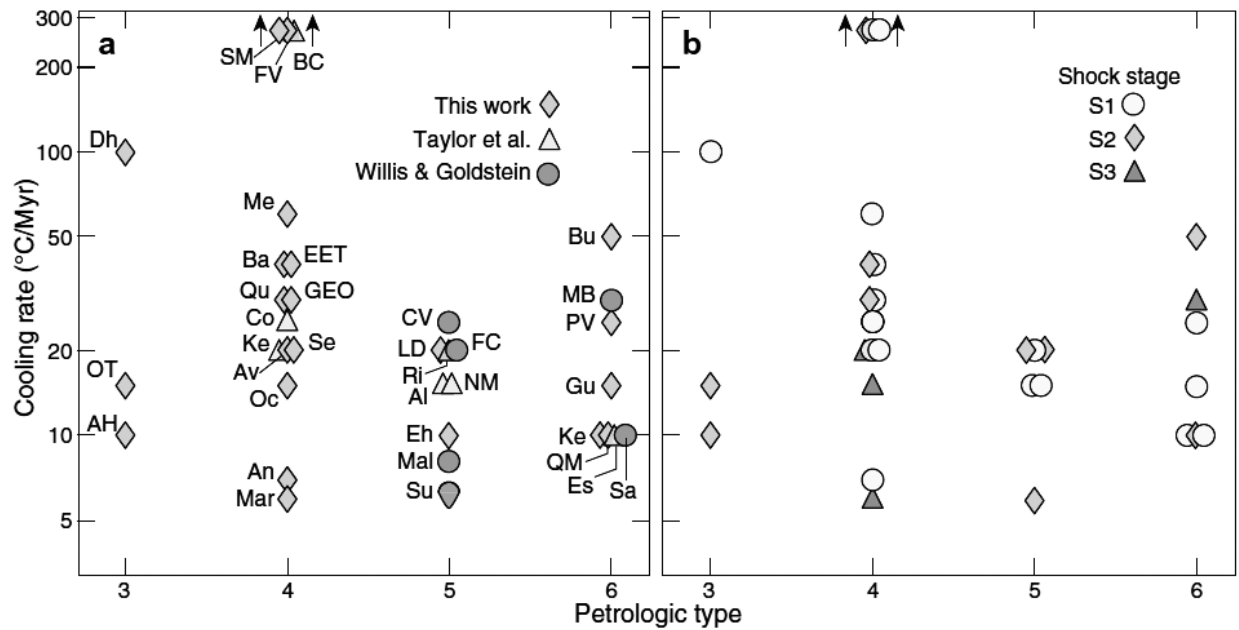

*Figure 2 (from Scott et al. 2014):* **Metallographic cooling rate vs. petrologic type for 35 H3–6 chondrites**: *(a) shows sources of data; (b) shows shock stage of chondrites. Cooling rate ranges in types 3–6 overlap considerably contrary to onion-shell model and are not correlated with shock stage. Most cooling rates have a precision of a factor of 2.*

Recently, *Scott et al.* (2014) have studied cloudy taenite, metallographic cooling rates, and shock effects in 30 H3–6 chondrites to elucidate the thermal and early impact history of the H chondrite parent body. They focused on H chondrites with old Ar–Ar ages (>4.4 Gyr) and unshocked and mildly shocked H chondrites, as strongly shocked chondrites with such old ages are very rare. They found that cooling rates for most H chondrites at 500 ºC are 10–50 ºC/Myr and do not decrease systematically with increasing petrologic type (see Figure 2) as predicted by the onion-shell model in which types 3–5 are arranged in concentric layers around a type 6 core. Some type 4 chondrites cooled slower than some type 6 chondrites and type 3 chondrites did not cool faster than other types, contrary to the onion-shell model. The three H4 chondrites that were used to develop the onion-shell model (*Trieloff et al.* 2003), Ste. Marguerite, Beaver Creek, and Forest Vale, cooled through 500 ºC at >5000 ºC/Myr, at least 50 times faster than any sample in a body that was heated by $^{26}$Al and cooled without impact disturbance. Since such fast-cooled H4 chondrites are rare and cooling rates of types 3–6 overlap considerably, *Scott et al.* (2014) infer that the suite of chondrites studied by *Trieloff et al.* (2003) are simply not representative of unreheated H4–6 chondrites. To explain their metallographic data, *Scott et al.* (2014) rely on recent work by *Ciesla et al.* (2013) and propose that several early impacts punctured the H chondrite parent body (bodies) to type 6 depths (assuming an onion-shell structure) while it was cooling, causing disturbances in the thermal histories of many H chondrites and leading to surfaces containing rocks that originated at a wide range of depths. They stress that the metallographic data do not require catastrophic disruption by impact during cooling. Considering the conflicting results reported above, it is reassuring that their results are coherent with the observed surface composition of large S-type asteroids (*Vernazza et al.* 2014; see section 4.2).

2.3.2 Constraints on the thermal evolution and impact history of the L chondrite parent body

About two thirds of L chondrite meteorites were heavily-shocked and degassed with $^{39}$Ar–$^{40}$Ar ages near 470 Myr (*Korochantseva et al.* 2007) suggesting that the L chondrite parent body suffered a major impact at ~470 Myr ago and catastrophically disrupted (see also *Heymann,* 1967; *Haack et al.* 1996 and the references therein). Metallographic cooling rates of lightly shocked or unshocked L chondrites are in the range 1–10ºC Myr$^{-1}$ (*Taylor et al.* 1987). These slow cooling rates imply that the original parent body had diameter D > 100 km (*Haack et al.* 1996).

It is striking that the timing of the shock event coincides with the stratigraphic age (467 ± 2 Myr) of the mid-Ordovician strata where abundant fossil L chondrites, meteorite-tracing chromite grains and iridium enrichment were found in the active marine limestone quarry in southern Sweden (*Schmitz et al.,* 1997, 2003, 2007; *Greenwood et al.,* 2007). The recent shocked and fossil L chondrites may thus apparently record the same event, a catastrophic disruption of a large main-belt asteroid that produced an initially intense meteorite bombardment of the Earth and at least ≈30% of the OC falls today.

2.3.3 Thermal history of the Itokawa parent body

On the basis of the mineralogy and mineral chemistry of ~50 Itokawa dust particles, we will briefly summarize the current understanding of the thermal history of the Itokawa parent asteroid, Itokawa likely being a collisional fragment from a once larger body (*Nakamura et al.* 2011). Mineralogical evidence indicates that Itokawa's parent body experienced thermal metamorphism and slow cooling (*Nakamura et al.* 2014). It reached a peak metamorphic temperature of ~800°C (*Nakamura et al.* 2011), which is lower by ~100°C than the peak temperature obtained for LL6 chondrites (*Slater-Reynolds and McSween,* 2005), suggesting that the most abundant material among Itokawa dust particles is likely LL5 material. In addition, the temperature during cooling appears to have remained higher than 700 °C for 7.6 Myr after CAI formation based on the absence of radiogenic $^{26}$Mg in plagioclase (*Yurimoto et al.* 2011a ; 2011b).

Both constraints have been used by *Wakita et al.* (2014) to reproduce the thermal evolution of the Itokawa parent asteroid via numerical simulations. They found that the Itokawa parent body appears to have formed with a radius greater than 20km and accreted between 1.9 and 2.2 Myr after CAI formation. Further modeling work by *Nakamura et al.* (2014) shows that a 50 km sized body formed around ~2.2 Myr after CAIs is one of the best formation models for the Itokawa parent asteroid. In short, both papers infer that the parent body was 20-50 km in radius or larger and that it accreted 1.9-2.2 Myr after CAI formation. However, since most LL chondrites and asteroids with LL chondrite spectral characteristics probably come from the large Flora family of asteroids, which dominates the inner part of the asteroid belt (section 5), it's likely that the Itokawa asteroid does too and is simply deficient in type 6 material. The parent body of the Flora family of asteroids was ~100 km in radius, judging from the inferred size of the known family members (*Durda et al.,* 2007).

## 3. Identification of the parent bodies of ordinary chondrites

In this section, we describe the outstanding efforts that have been accomplished by a large fraction of the asteroid-meteorite community to unambiguously identify the parent bodies of ordinary chondrites. These efforts include four decades of telescopic observations, more than 20 years of laboratory measurements and experiments and the first asteroid sample return mission.

3.1. Ordinary chondrite parent bodies among S-type asteroids?

The parent bodies of ordinary chondrites (OCs) have been searched for via visible and near-infrared photometric and/or spectroscopic observations for more than 40 years. Early photometry of asteroids based on photoelectric sensors (*Chapman et al.* 1971, *Hapke* 1971) revealed two predominant colorimetric groups: (a) those with slightly reddish sloping reflectance spectra throughout the visible out to 1 micron and with an absorption feature starting longward of 0.7 micron and with a band minimum around 0.95 microns (such asteroids are commonly in the inner half of the belt) and (b) those with flat reflectance spectra (commonly located in the outer half of the main belt – *Chapman,* 2004). *Chapman et al.* (1975) introduced the taxonomy that called the reddish, moderate-albedo (a) group the "S-type" (mnemonic for silicaceous because the 0.95 micron band had been interpreted as being due to silicate minerals) and the neutral-colored, low albedo (b) group the "C-type" (mnemonic for carbonaceous, by analogy with carbonaceous meteorites). Thus began the C-, S- and, later, M-... asteroid taxonomy (*Chapman* 2004), which has now been expanded to include most letters of the alphabet and has been further refined by many researchers (the latest taxonomy is due to *DeMeo et al.* 2009; see *DeMeo et al.* chapter for a review). More recent surveys including the SDSS one that surveyed more than 100,000 asteroids in the visible (*DeMeo and Carry* 2013, 2014) have not changed the global view presented in *Chapman et al.* (1975), that is, most asteroids can still be regarded as being S or C.

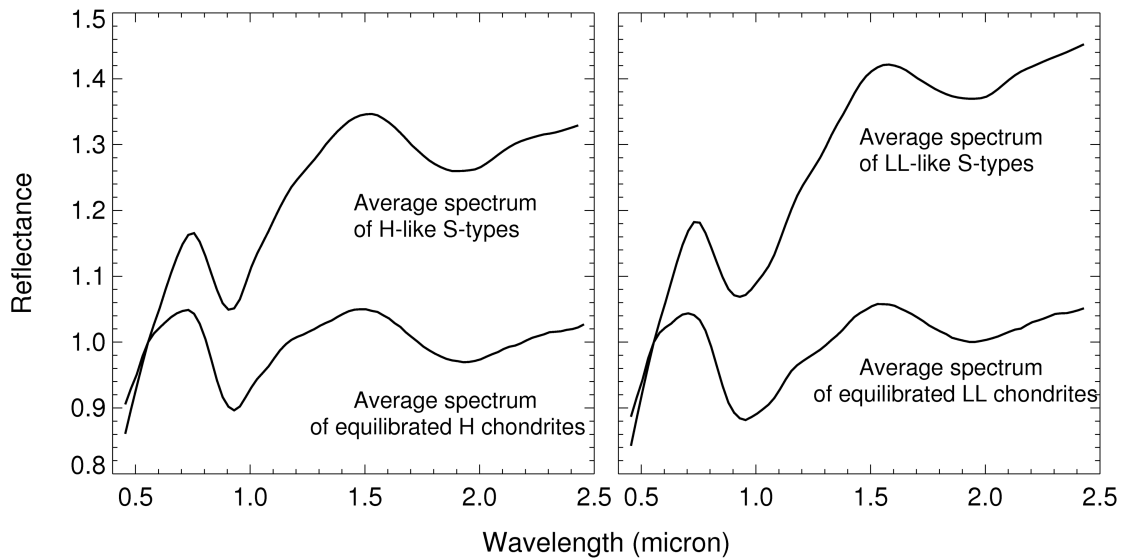

*Figure 3 (adapted from Vernazza et al. 2014):* **Spectral comparisons of S-type asteroids and ordinary chondrite meteorites.** *Comparison between the visible to near-infrared spectral signatures of main belt S-type asteroids (left: asteroids with H-like compositions; right: asteroids with LL-like compositions) and the average spectra of equilibrated (type 4 to 6) H and LL chondrite meteorites. Space weathering processes similar to those acting on the Moon (e.g.,*

*Pieters et al. 2000)* redden the ordinary chondrite like spectrum of a fresh asteroid surface, giving it the appearance of an S-type spectrum (e.g., Vernazza et al. 2008 and references therein).

Early comparisons between meteorite and asteroid optical reflectance spectra (e.g., *Chapman and Salisbury* 1973) suggested that only S-type asteroids have spectral characteristics (e.g., absorption band near 0.95 micron) compatible with those seen in OCs although S-type asteroids have a redder reflectance. Subsequent measurements in the near-infrared range have not altered this picture (e.g., *Gaffey et al.* 1993, *Vernazza et al.* 2008, *Vernazza et al.* 2009, *DeLeon et al.* 2010, *Thomas et al.* 2010, *Dunn et al.* 2013, *Vernazza et al.* 2014). The spectral slope difference between OC and S-type spectra (see Figure 3) has, however, been an obstacle for more than 30 years to establishing a definite link between both groups and has constituted one of the greatest conundrums in the asteroid-meteorite field (see *Chapman* (2004) for an extensive review on this subject). The conundrum was even reinforced on the basis of mineralogical arguments (*Gaffey et al.* 1993) that were suggesting that a large fraction of S-type asteroids are partly differentiated and thus not OC-like (*Abell et al.* 2007). Such view was held by a small fraction of the community until the Itokawa sample return (*Abell et al.* 2007).

3.2 Hayabusa confirms a linkage between S-type asteroids and ordinary chondrites

The Japanese spacecraft Hayabusa stored in the M-V-5 rocket was launched on May 9th 2003. The aims of the Hayabusa mission were (1) to confirm a linkage between S-type asteroids and ordinary chondrites, and (2) to uncover the mechanisms of space weathering, responsible for the mismatch of reflectance spectra between S-type asteroids and ordinary chondrites (see chapter by *Yoshikawa et al.* for an overview of the mission and its main results). Hayabusa arrived at the S-type asteroid Itokawa (Fig. 4) in September 2005 and observed its surface by a variety of on-board apparatus (*Fujiwara et al.* 2006). The results of the in-situ measurements indicated that the reflectance spectra (*Abe et al.* 2005) are similar to those of equilibrated LL chondrites thus confirming *Binzel et al.* (2001)'s earlier prediction. Hayabusa touched down twice on smooth terrain "MUSES-C Regio" (Fig. 4a) and tried to recover surface samples from the asteroid. Although projectiles that planned to hit the asteroid surface did not fire, the sampling horn of the spacecraft touched the surface several times which probably stirred up Itokawa particles into the spacecraft (*Yano et al.* 2006). The spacecraft finally came back to Earth in June 2010, five years after the touchdown.

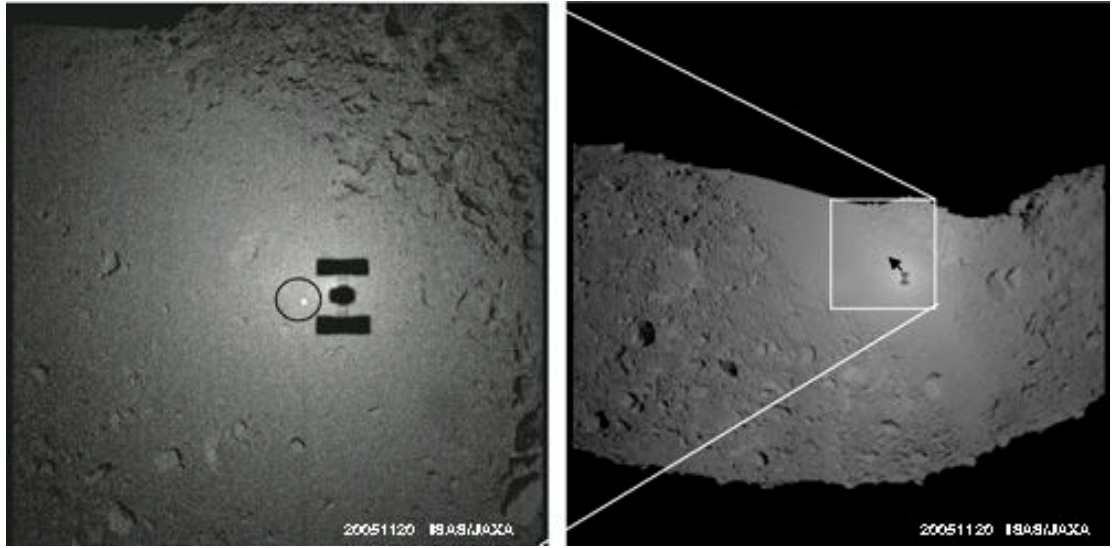

*Figure 4: **Itokawa's landing site.** (A) Hayabusa is going down to MUSES C Regio and the shadow of the spacecraft is projected on the surface. (B) A whole view of MUSES-C region. JAXA digital images P-043-15285.*

Numerous small particles mostly less than 50μm in size were found in the Hayabusa sample container and were recovered one by one by electrostatic manipulator operation in a pure nitrogen atmosphere (*Nakamura et al.* 2011; *Yada et al.* 2014). The initial analysis for basic characterization of approximately 60 Itokawa dust particles showed that the mineralogy and mineral chemistry of the Itokawa dust particles are identical to those of equilibrated LL chondrites (*Nakamura et al.* 2011; see Figure 5). Oxygen isotope ratios of silicates also indicate the similarity to equilibrated LL chondrites (*Yurimoto et al.* 2011; *Nakashima et al.* 2014). Minor and major elemental abundances were found to be similar to those of ordinary chondrites (*Ebihara et al.* 2011). In addition, the modal abundance of minerals, bulk density, porosity, and grains size of the Itokawa particles were found to be similar to those of LL chondrites (*Tsuchiyama et al.* 2014).

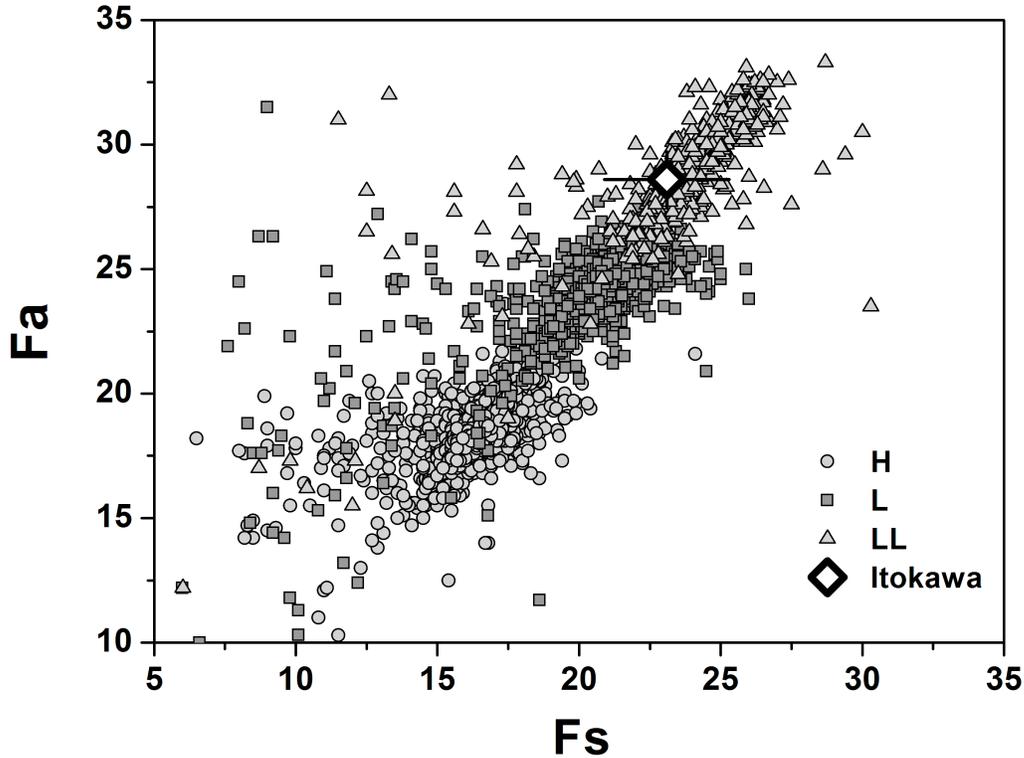

*Figure 5 (Figure from Nakamura et al. 2011):* ***Comparison between the mineral chemistry of the Itokawa particles and that of H, L, and LL chondrites.***

Figures 6a and 6b show typical Itokawa particles. Olivine is the most abundant mineral as is the case for LL chondrites. Most of the olivine particles have compositions equilibrated around $Fa_{29}$ (Figure 5), which is within the compositional range of equilibrated LL chondrites, but some olivines show Mg-rich compositions, suggesting that they were on the way to final equilibration (*Nakamura et al.* 2011). The next most abundant minerals are, as usually observed among equilibrated LL chondrites, low- and high-Ca pyroxenes and plagioclase and their compositions are also almost homogeneous.

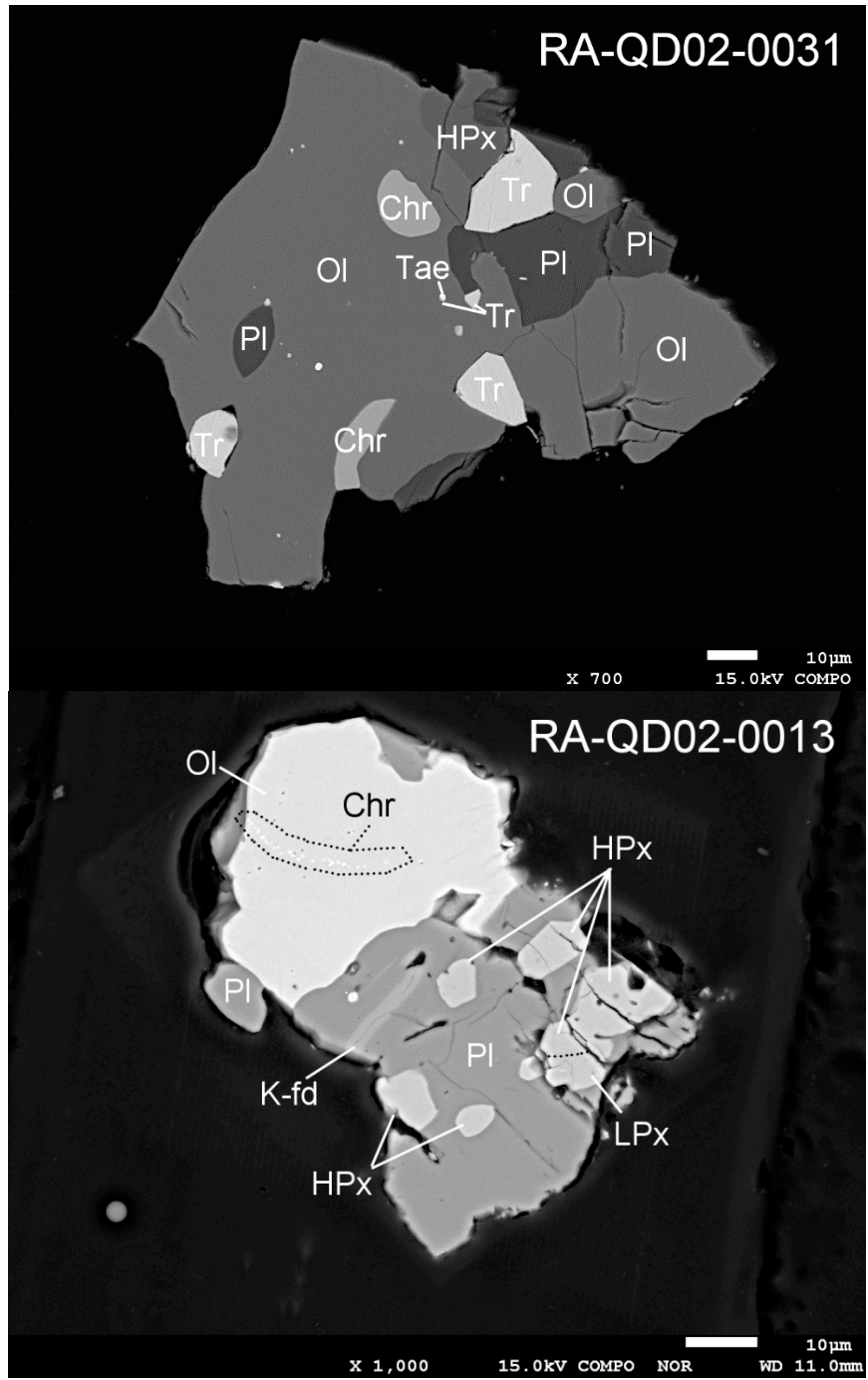

*Figure 6: **Back-scattered electron images of polished sections of typical Itokawa particles** (Top: RA-QD02-0031; Bottom: RA-QD02-0013). Abbreviation: Ol=olivine, Pl=plagioclase, Tae=taenite, Tr=troilite, Chr=chromite, LPx=low-Ca pyroxene, HPx=high-Ca pyroxene, K-fd=K feldspar*

In summary, all measurements performed on Itokawa both in situ and in terrestrial laboratories indicate that the asteroid has an LL-like composition. This result confirms what many researchers had predicted (e.g., *Pellas* 1988), namely that S-type asteroids comprise the parent bodies of ordinary chondrites. This fact implies that space weathering processes are generally responsible for the spectral mismatch between OCs and S-type asteroids (e.g., *Sasaki et al.* 2001, *Strazzulla et al.* 2005; see chapter by *Brunetto et al.* for a detailed review), which we summarize in the next subsection.

3.3 Space weathering processes responsible for the spectral mismatch between OCs and S-type asteroids

Over the last 15 years, laboratory experiments simulating space weathering effects on OCs and its main minerals (olivine, pyroxene) (e.g., *Sasaki et al.* 2001, *Strazzulla et al.* 2005, *Marchi et al.* 2005, *Brunetto et al.* 2006; see chapter by *Brunetto et al.* for more information) have established unambiguously that OC-like material naturally reddens (spectrally) in space under both the bombardment by micrometeorites and irradiation by solar wind particles, a conclusion that was reinforced by the analysis of both the images and near-infrared spectra of Itokawa's surface collected by Hayabusa as well as the S-type samples brought back to Earth by Hayabusa (*Nakamura et al.* 2011).

Spatially resolved images and near-infrared spectra of Itokawa's surface by Hayabusa (*Hiroi et al.* 2006) revealed developing space weathering on Itokawa's surface. Specificaly, *Hiroi et al.* (2006) showed that dark areas on Itokawa possess on average redder spectra than bright areas in agreement with space weathering spectral trends predicted by laboratory experiments (e.g., *Sasaki et al.* 2001, *Strazzulla et al.* 2005; see chapter by *Brunetto et al.*).

Subsequent laboratory measurements of the three-dimensional structure of the Itokawa particles determined by X-ray microtomography suggested that they have been impacted by meteoroids (*Tsuchiyama et al.* 2011); the discovery of micro-craters on the surface of some Itokawa particles reinforced the evidence of interplanetary dust particles impacting asteroid surfaces (*Nakamura et al.* 2012). Large amounts of solar wind noble gases were also detected within the Itokawa particles, implying that solar-wind implantation into the regolith particles is a common process (*Nagao et al.* 2011). In parallel, a careful inspection of the Itokawa surface samples revealed sulfur-bearing and sulfur-free Fe-rich nanoparticles at the outermost layers of the Itokawa particles (*Noguchi et al.* 2011), thus identifying the reddening agent responsible for the color difference between S-types and OCs. The results of further TEM analysis identified three types of particle surface modifications induced by space weathering processes (*Noguchi et al.* 2014). In summary, all these results indicate that the Itokawa dust particles - recovered from the space-weathered area MUSES-C Regio - record intensive interactions between the asteroid regolith and incoming particles such as solar wind ions and interplanetary dust. The Hayabusa mission has thus confirmed what was suspected for a long time, namely that OCs originates from S-type asteroids and that the spectral mismatch between OCs and S-types is mainly due to space weathering processes.

As a consequence (and fully ironically considering how long the S-type/OC conundrum did last), the *normal* appearance of an OC-like asteroid is an S-type asteroid (spectral mismatch between OCs and their parent bodies), while the *anomalous* appearance of an OC-like asteroid is a Q-type asteroid (spectral match between OCs and their parent bodies)! The existence of Q-type asteroids, which indicates that rejuvenating processes do occur on some asteroids (see chapter by *Brunetto et al.*), has offered us the opportunity to unveil a new physical process in asteroid science, namely seismic shaking during planetary encounters (*Nesvorny et al.* 2005, *Marchi et al.* 2006, *Binzel et al.* 2010, *DeMeo et al.* 2014, and chapter by *Binzel et al.*). Note that thermal fatigue (*Delbo et al.* 2014) and spin-up due to the YORP effect (see chapter by *Vokrouhlicky et al.* and references therein) may also play a role in refreshing asteroid surfaces.

3.4. How many S-type asteroids actually have ordinary chondrite-like mineralogies?

At a time where it was still not fully accepted that S-type asteroids comprise the parent bodies of OCs, researchers started comparing the surface mineralogy of S-type asteroids with that of OCs to test a possible link between the two groups of objects (e.g., *Gaffey et al.* 1993). Such compositional investigation could not be performed in the visible domain only: it required extended wavelength coverage into the near infrared (ideally a spectral coverage over the 0.4-2.5 micron range; see chapter by *Reddy et al.*). In this extended wavelength range, both S-types and OCs possess prominent 1 and 2 micron spectral features due to the presence of olivine and pyroxene in the two populations (see Figure 6).

*Gaffey et al.* (1993) were the first to perform this type of mineralogical investigation for a large number of S-type asteroids (39 objects; it actually turns out that only 32 of these objects are S-type main belt asteroids). They used a classification scheme (see Figure 7) based on Band Area Ratios (BAR; Area of Band II divided by Area of Band I) and Band I centers to break the S-types into seven subtypes [from S(I) to S(VII)], corresponding to varying ratios of olivine to pyroxene. The BAR value tends to increase and the Band I center value tends to decrease as the subtype number increases in value, which indicates an increasing pyroxene concentration (*Burbine* 2014). Their S(I) type has an olivine-dominated mineralogy and in the new classification by *DeMeo et al.* (2009) corresponds mostly to A-types (and thus not to S-types anymore) while the S(IV) subtype has an olivine-pyroxene mineralogy similar to that of ordinary chondrites. Both the S(II) and S(III) subtypes were interpreted as having olivine-dominated mineralogies with a high-Ca pyroxene component, while the S(V) and S(VII) were interpreted as having a higher and significant high-Ca pyroxene component.
The main conclusion of *Gaffey et al.* (1993)'s investigation is that few S-type asteroids have silicate mineralogies consistent with those of ordinary chondrites and that the diversity within the S-class arises from several sources, including the coexistence of undifferentiated, partially differentiated, and fully differentiated bodies within the general S-type population.

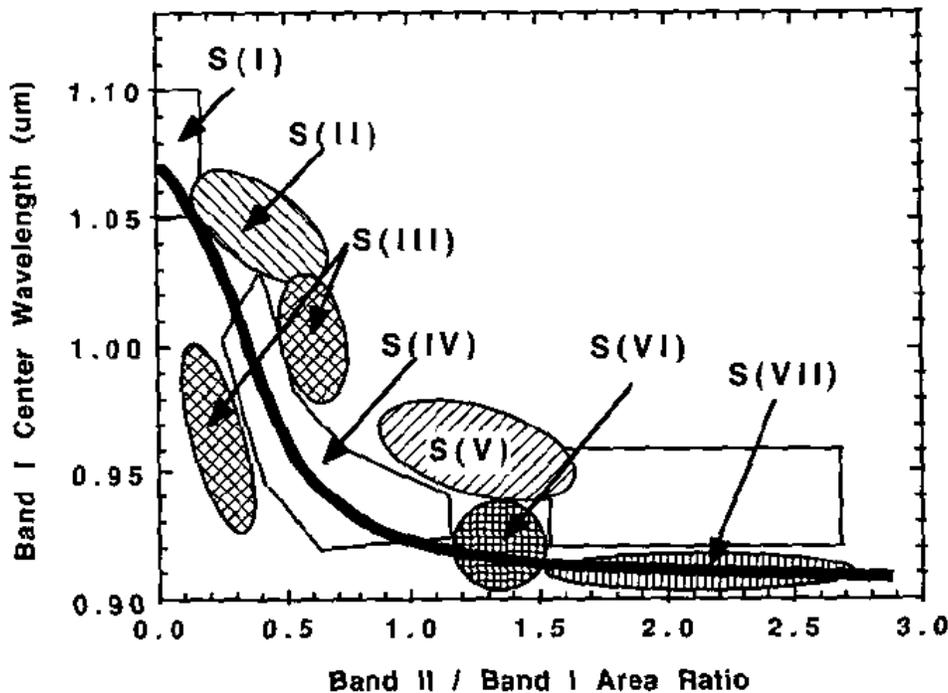

*Figure 7 (from Gaffey et al. 1993):* **Distribution of Gaffey et al. (1993) S subtypes on a plot of Band Area Ratio versus Band I center**. *The thick black line is the olivine–orthopyroxene mixing line. The ordinary chondrite region is the same as the S(IV) region. The HED region (rectangle above SVII) is also plotted.*

Concerning the few objects that were considered ordinary chondrite-like, *Gaffey et al.* (1993) noticed that they are concentrated near the 3:1 Kirkwood gap at 2.5 AU and proposed that their favorable location close to this resonance might explain why ordinary chondrites are so abundant among meteorite falls. A few years later, *Gaffey and Gilbert* (1998) proposed asteroid (6) Hebe as the probable parent body of H-type ordinary chondrites and, since then, many authors have relied on this association in their studies (e.g., *Akridge et al.* 1998, *Ghosh et al.* 2003, *Bottke et al.* 2010 abstract, *Henke et al.* 2012, 2013). In the early 2000s, the idea that several S-type asteroids have high fractions of high-Ca pyroxene on their surfaces, which is an indication that these bodies have undergone either melting or partial melting and that they can thus not be linked to OCs, was brought up again by *Sunshine et al.* (2004), *Hardersen et al.* (2006) and *Abell et al.* (2007) in the case of Itokawa.

During the following years, the emergence of numerous high quality spectroscopic measurements in the near-infrared (obtained essentially with the NASA IRTF) along the growing utilization within the community of new spectral analysis tools (Modified Gaussian Method (MGM), radiative transfer model; see chapter by *Reddy et al.*) has progressively overturned the view by *Gaffey et al.* (1993) that many S-types are compositionally unconnected to OCs. Using both MGM and a radiative transfer model to constrain the mineralogy of a large sample of S-type NEAs, *Vernazza et al.* (2008) showed that most of these objects have silicate mineralogies that are compatible with those of OCs. Their findings were later confirmed by *DeLeon et al.* (2010), *Dunn et al.* (2013) and *Thomas et al.* (2014) for even larger samples of S-type near-Earth asteroids and by *Vernazza et al.* (2009), *De Leon et al.* (2010) and *Vernazza et al.* (2014) for large samples of S-type main belt asteroids. Various authors derived the mineralogy of smaller samples of S-types and found OC-like mineralogies in most cases (*Binzel et al.* 2001, *Binzel et al.* 2009, *Reddy et al.* 2009, *Reddy et al.* 2011a, *Reddy et al.* 2011b, *Fieber-Beyer & Gaffey* 2011, *Reddy et al.* 2012, *Fieber-Beyer et al.* 2012, *Gietzen et al.* 2012, *Sanchez et al.* 2013, *Fieber-Beyer & Gaffey* 2014).

3.5 Are there meteorites other than the ordinary chondrites derived from S type asteroids?

S-type asteroids are a diverse group of objects and one cannot exclude that some meteorites other than the ordinary chondrites could also originate from them (e.g., *Burbine et al.,* 2003). Possibilities include differentiated meteorites, as proposed by *Sunshine et al.* (2004), such as pallasites, brachinites, ureilites, lodranites, winonaites, IAB irons, and mesosiderites (*Gaffey et al.,* 1993).

Primitive achondrites (including winonaites, lodranites, acapulcoites, and brachinites) are meteorites with affinities to chondrites. They have undergone partial melting but the silicate and metal portions have not segregated, and so they retain their chondritic bulk composition; they are therefore likely to have spectral similarities to chondrites. Indeed, it appears that both lodranites and acapulcoites have spectral properties and ol/(ol+ low Ca-px) ratios that are very similar to OCs and H chondrites in particular (see Figure 8) although they are, on average, slightly more pyroxene-rich than H chondrites. However, there are two reasons as to why most H-like S-type asteroids may not be linked to either of these meteorite classes. The first is that the center of band II for both lodranites and acapulcoites is located between 1.8 and 1.9 micron while the band center for H chondrites (and H-like S-types; see *Vernazza et al.* 2014) and other OCs is located between 1.9 and 2 microns indicating differences in terms of pyroxene composition between the two groups. The second reason is that both lodranites and acapulcoites represent together ~0.2% of the falls, implying that they are ~150 times less abundant than H chondrites among falls (and ~400 times less abundant than OCs in general). Although it is clear that even at similar densities and tensile strength (which is typically the case for the H, L, and LL chondrites but also for lodranites and acapulcoites) meteorite falls are not exactly representative of the compositional diversity of the asteroid belt (*Vernazza et al.* 2014), they still are within a factor of ~10, implying that lodranite-like and acapulcoite-like bodies should be rare among S-types.

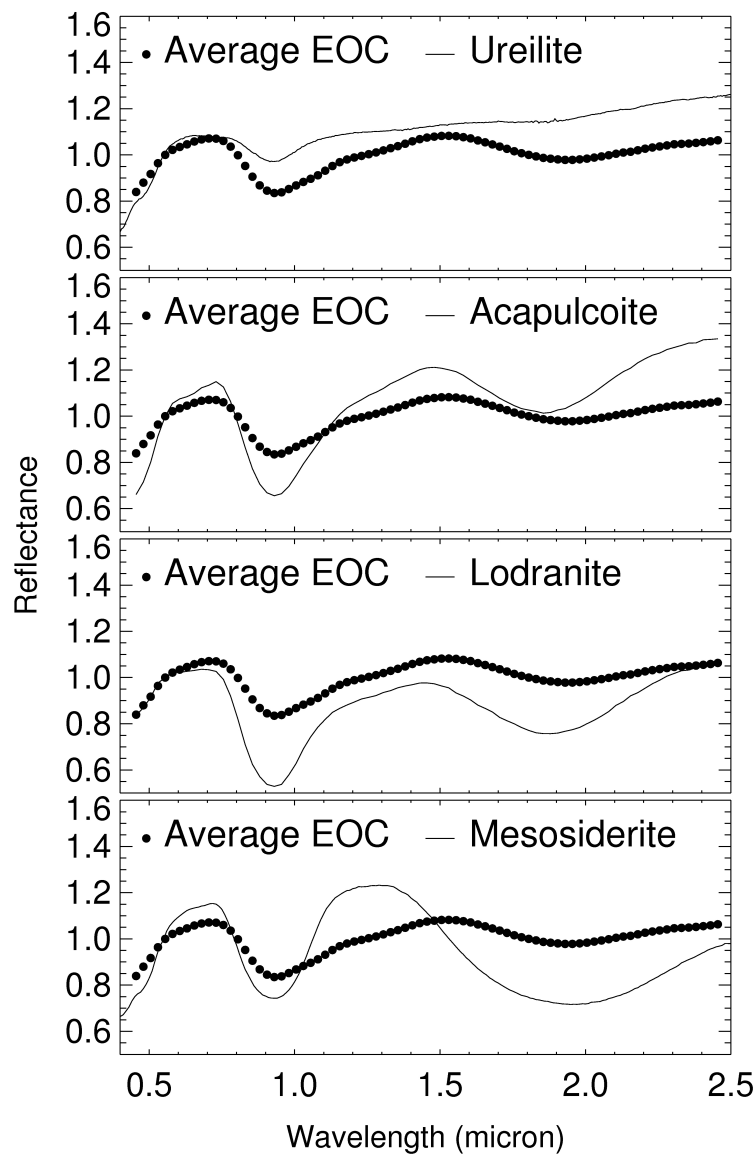

*Figure 8:* **Comparison between an average spectrum of equilibrated H, L and LL chondrite and the spectra of ureilites, acapulcoites, lodranites and mesosiderites** (data taken from the RELAB database). *While the spectra of both ureilites and mesosiderites appear quite different*

*from the OC spectra, both lodranites and acapulcoites have similar spectra with respect to OCs.*

Mesosiderites are complex brecciated meteorites consisting of metal rich and silicate rich portions. Their spectra (*Burbine* 2007) are at odds with those of S-type asteroids and/or OCs (see Figure 8). The same applies to ureilites which are igneous meteorites composed of olivine and pyroxene with notably large amounts of carbon (several percent). The possibility that ureilites could have an S type asteroid source could be tested when the Almahatta Sitta ureilite breccia fell near Nahr an Nil in Sudan in 2008. The spectroscopic features of the parent asteroid for Almathatta Sitta, 2008 TC3, were determined by *Jenniskens et al.* (2010). They concluded that the asteroid most closely resembled an F type asteroid, and that this asteroid type is more likely to be the parent to ureilites than S type asteroids.

Brachinites whose spectra are not displayed here are far more olivine-rich than LLs. Their spectra are great analogs for A-type asteroids but certainly not for S-type ones (*Sunshine et al.* 2007). Finally, pallasites are also olivine-rich meteorites. As in the case of brachinites, their parent asteroids should be hidden among A-type asteroids (*Sunshine et al.* 2007).

## 4. Constraints on the formation and evolution of ordinary chondrite parent bodies from telescope observations

The ability to link specific ordinary chondrite classes (H, L and LL) to specific S-type asteroids via visible and near-infrared spectroscopy has not only provided us with a better understanding of the formation mechanism of OC parent bodies and planetesimals in general, it has also boosted our understanding of asteroid dynamics and in particular of the delivery process of meteorites and NEAs from the main belt to the near Earth space.

4.1 What have we learned from S-type NEAs?

Spectroscopic observations of more than 400 near-Earth asteroids (NEAs) in visible wavelengths show that 65% of NEAs have S- and Q-type spectral properties (*Binzel et al.* 2004). When corrected for discovery biases (*Stuart & Binzel* 2004), the near-Earth population of S- and Q-type asteroids is estimated to be 36% of the total NEA population. *Vernazza et al.* (2008) used both MGM (Modified Gaussian Modeling) and a radiative transfer model to analyze the visible and near-infrared spectra of 38 S- and Q-type NEAs from *Binzel et al.* (2004) and reported that most NEAs (~2/3) have spectral properties similar to LL chondrites (see Figure 9). This result is surprising, because LL chondrites are the least abundant ordinary chondrites (they represent only 10% of all ordinary chondrites, and 8% of all meteorites). *Vernazza et al.* (2008) argued that the NEAs we sample telescopically (with radii of 0.3 km to 10 km) are not the immediate parent bodies of smaller objects that fall to Earth as meteorites (that is, pre-atmospheric meteorite parent bodies having radii on the order of metres) and that different dynamical mechanisms (mainly a size-dependent process such as the Yarkovsky effect) and/or main-belt source regions may be responsible for supplying these two sample populations. They further proposed the Flora family (near the $\nu_6$ secular resonance) as the source region for NEAs with LL-like compositions.

*Thomas and Binzel* (2010) used MGM to determine meteorite analogues for a sampling of S-type NEAs and then used the *Bottke et al.* (2002) dynamical model to determine the probable source region for each of those NEAs. They found that H chondrites have a higher than average delivery preference though the 3:1 mean motion resonance whereas LL chondrites are mostly injected in the near-Earth space via the ν6 resonance. *De León et al.* (2010) and *Dunn et al.* (2013) also concluded that the Flora family is the dominant source of NEAs and LL chondrites, based on inferred mineralogies of large samples of NEAs. A more recent study by *Thomas et al.* (2014) of an even larger sample of S-type NEAs (109 objects) concluded that NEAs have elevated percentages of potential LL ordinary chondrites compared to the meteorite fall statistics. However, they also concluded that that the relative proportions among NEAs of the ordinary chondrite spectral types lie somewhere between the H and L chondrite dominant meteorite fall statistics and the LL chondrite spectral types dominant NEA statistics as calculated by *Dunn et al.* (2013) and seen in *Vernazza et al.* (2008) and *De León et al.* (2010).

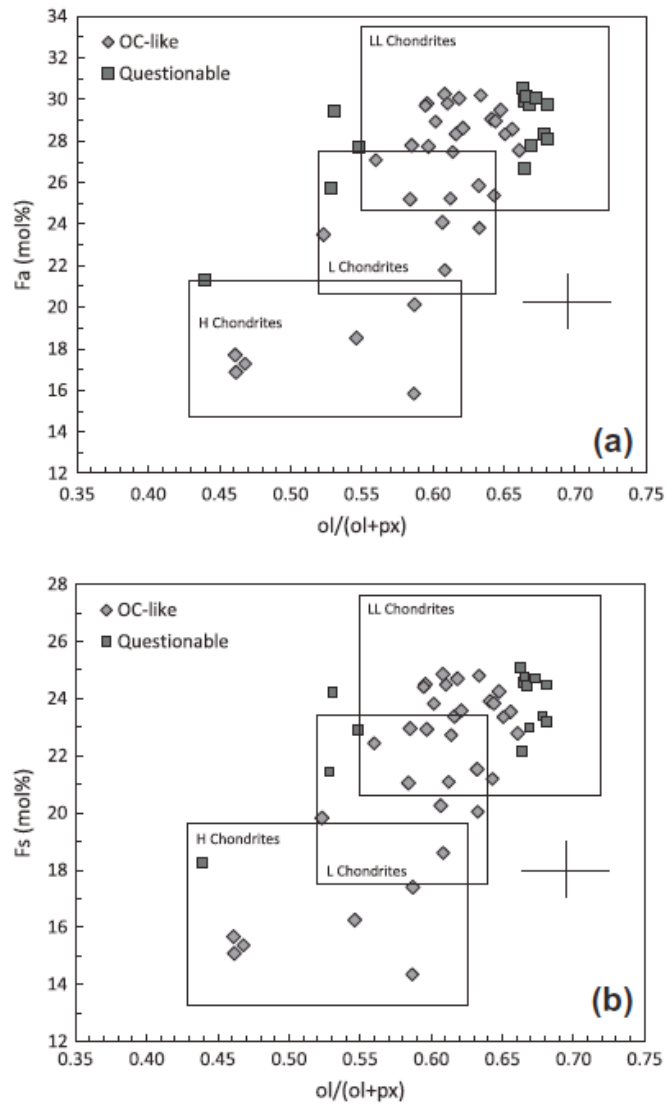

*Figure 9 (from Dunn et al. 2013):* **Comparison between the mineralogies of 47 NEAs and those of ordinary chondrites.** *NEAs are plotted as (a) ol/(ol + px) vs. mol% Fa in olivine and (b) ol/(ol + px) vs. mol% Fs in low-Ca pyroxene. Compositional regions for the H, L, and LL chondrites were defined by Dunn et al. (2010). NEAs with definitive ordinary chondrite-like spectral parameters are represented by boxes, while NEAs with questionable spectral classification are*

*represented by diamonds. Error bars represent the least mean square error of spectrally derived mineralogies: 0.03 for ol/ol + px, 1.3 mol% Fa, and 1.4 mol% Fs.*

4.2 What have we learned from S-type main belt asteroids (MBAs)?

As stated above, *Gaffey et al.* (1993) observed 32 main belt S-types and found that most of these objects do not have OC-like mineralogies. They suggested that most of the non OC-like S-type asteroids originate from partially or fully differentiated bodies. More recently, *Vernazza et al.* (2014) revisited this question by conducting an extensive spectroscopic survey of 83 main belt S-type asteroids and 3 S-type families, and obtaining the biggest spectral data set yet assembled for the largest S-type asteroids (95% of all objects larger than 60km). In parallel, they built up the existing database of ordinary chondrite laboratory spectral measurements by collecting spectra for 53 unequilibrated ordinary chondrites (i.e., for the most primitive OCs). Thus the database now spans a much broader range of temperature history (from unheated to significantly metamorphosed) than previously analyzed. On the basis of these two spectral surveys, they reach the following conclusions:
1) most S-type asteroids, including large ones (D≈100-200 km), though not members of one family, are distributed into two well-defined compositional groups (see Figure 10), Hebe-like and Flora-like (H-like and LL-like). This indicates that identical compositions among multiple asteroids are a natural outcome of planetesimal formation, making it possible that meteorites within a given class originate from multiple parent bodies;
2) the surfaces of nearly all of these asteroids (up to 200 km) show the same compositional characteristics as high temperature meteorites (type 4 to 6 OCs) that were metamorphosed in their interiors, these exposed interiors being a likely consequence of impacts by small asteroids (D<10km) in the early solar system (*Ciesla et al.* 2013);
3) the lack of compositional variation within both H-like asteroid families and the surfaces of the family members showing the same compositional characteristics as high temperature meteorites (type 4 to 6 OCs) is consistent with their parent bodies having been metamorphosed throughout, which implies – following current thermal models - that their formation process must have been rapid. Note that such a short duration of accretion as implied by the observations of *Vernazza et al.* (2014) is consistent with current models of planet formation through streaming and gravitational instabilities (*Youdin & Goodman* 2005, *Johansen et al.* 2007, *Chiang & Youdin* 2010) that show that bodies of several hundred kilometers in size form on the time-scale of a few orbits (*Johansen et al.* 2011, *Youdin* 2011, *Johansen et al.* 2012);
4) LL-like bodies formed closer to the Sun than H-like bodies, a possible consequence of radial mixing and size sorting of chondrules in the protoplanetary disk prior to accretion;
5) LL-like bodies formed on average with larger sizes than H-like bodies.

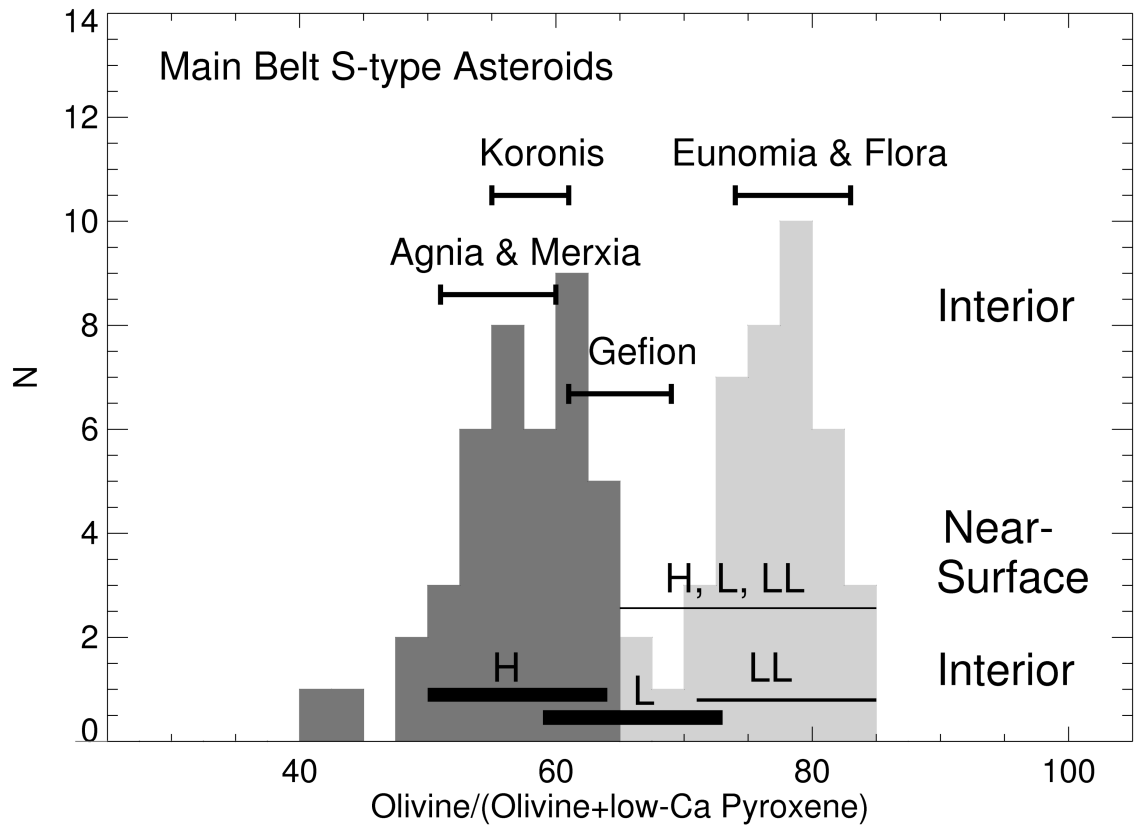

*Figure 10 (from Vernazza et al. 2014):* **Bimodal compositional distribution of main-belt S-type asteroids.** *The S-type sample comprises 83 objects, including 54 out of the 56 main-belt S-types with D>60km. Objects belonging to collisional families are not included in the histogram counts. Instead, the compositional range for the 6 main asteroid families (Agnia, Merxia, Koronis, Gefion, Eunomia, Flora) are shown at the top. The compositional ranges for the individual ordinary chondrite classes of « Interior » samples (H, L and LL having petrologic types >3.5; temperature histories >400°C) are shown in red. The blue line denotes the compositional range for the least metamorphosed OCs (Types 3.0-3.5; temperatures <400°C) that are interpreted as surface samples. The thickness of the various compositional ranges for meteorites is proportional to their fall statistics. Finally, the diversity of S-type asteroids includes compositions outside the range of ordinary chondrites as seen for the two objects having ol/(ol + low-Ca px) < 45%.*

## 5. Summary and Conclusions

The nature of the parent bodies of ordinary chondrites is no longer a mystery (*Nakamura et al.* 2011). It has now been clearly established that the parent bodies of ordinary chondrites are hidden among the S-type asteroids. Actually, most S-type asteroids are plausible OC parent bodies as they share similar spectral properties and thus similar mineralogies (*Vernazza et al.* 2014).

Spectroscopic surveys have allowed determining rather precisely the compositional distribution among S-type NEAs and S-type MBAs. It appears that the majority of the S-type NEAs have LL-like mineralogies implying important compositional differences between NEAs and meteorites (*Vernazza et al.* 2008, *De León et al.* 2010, *Dunn et al.* 2013, *Thomas et al.* 2014), and thus different source regions for both populations. Concerning S-type MBAs, it appears that most S-type asteroids, including large ones (D≈100-200 km), are distributed into two well-defined compositional groups, Hebe-like and Flora-like (H-like and LL-like), with Flora-like bodies being located - on average – closer to the Sun than Hebe-like bodies (*Vernazza et al.* 2014). In addition, the surfaces of nearly all of these asteroids (up to 200 km) show the same compositional characteristics as high temperature meteorites (type 4 to 6 OCs) that were metamorphosed in their interiors suggesting that impacts may have played an important role in their structural evolution (*Vernazza et al.* 2014). Interestingly, the metallographic cooling rates of H chondrites are in agreement with those observations as they suggest that several early impacts punctured the H chondrite parent body (bodies) while it was cooling, causing disturbances in the thermal histories of many H chondrites and leading to surfaces containing rocks that originated at a wide range of depths (*Taylor et al.* 1987; *Scott et al.* 2011, 2013, 2014).

Although significant progress has been made since the Asteroid III book in linking OCs to S-types, there are several issues remaining that require critical new data:

a) The source regions of the individual OC classes remain to be determined. Currently, we have only predictions for the source regions of both L and LL chondrites, namely the Gefion family is predicted, based on dynamical and compositional aspects (*Nesvorny et al.* 2009, *Vernazza et al.* 2014), to be the source of L chondrites while the Flora family is proposed to be the source of most LL chondrites (e.g., *Vernazza et al.* 2008). The source of H chondrites is currently unknown; *Vernazza et al.* (2014) have shown that several prominent S-type families could be the source of these meteorites.
Over the next decade, the ongoing development and installation of several fireball observation networks across the world (e.g., FRIPON: http://ceres.geol.u-psud.fr/fripon/) will help solve this long-standing issue. These camera networks will observe a statistically significant number of falls within the next 10 years and thus allow the determination of precise orbits (and hence directly the source region) for a large number of meteoroids. Hopefully, they will also allow us to recover a non-negligible number of meteorites. Recent successful meteorite recoveries include the Tagish Lake, Almahata Sitta, Maribo, Kosice, Bunburra Rockhole and Sutter's Mill falls (see chapters by Borovicka et al. and Jenniskens et al.).

b) It remains to be explained why there are so few LL chondrites and so many H and L chondrites among falls (*Vernazza et al.* 2008) given that the prominent LL-like Flora family is located next to the most prolific (in terms of delivery) resonance (namely $\nu_6$). In other terms, the observed compositional difference between meteorites and NEAs (Vernazza et al. 2008) remains to be explained.

c) It remains to be established whether the individual OC classes (H, L, LL) originate from one or several parent bodies. Spectroscopic surveys have opened the possibility that the individual OC classes can come from several parent bodies (*Vernazza et al.* 2014). A new generation of meteorite measurements may shed light on this issue. Fireball observation networks will also help progress on this question.

d) The diversity of the fall statistics between H, L, and LL chondrites as a function of petrologic type (62% of H chondrites are type 4 and 5, while type 6 represents only 21%, 68% of Ls are type 5 and 6 and 59% of LLs are type 5 and 6; *Hutchison* 2004) remains to be explained. Modeling the thermal evolution of ordinary chondrite parent bodies as a function of the parent body's size may certainly help shedding light on this issue. (Hebe not necessarily being the parent body of H chondrites, a D~100 km sized parent body should be envisioned along with a D~200 km sized one for H chondrites, while larger parent bodies should be considered for the L and LL chondrite parent bodies ; see *Vernazza et al.* 2014). Determining whether all meteorites within a given ordinary chondrite class (e.g., H chondrites) come from the same parent body via both fireball observation networks and refined measurements on OCs will also help explaining the diversity of the fall statistics.

e) Spectroscopic surveys have revealed that the parent bodies of LL chondrites formed on average larger than those of H chondrites (*Vernazza et al.* 2014). An explanation of this observation remains to be found.

f) The parent bodies of lodranites and acapulcoites are likely hidden among S-type asteroids. A careful spectral analysis of these meteorites will help to pinpoint their possible parent bodies.

g) The nature of the parent bodies of HH (Burnwell and Willaroy), H/L and L/LL chondrites remains to be understood. The paucity of H/L and L/LL chondrites may hint at a rather clear separation between the H, L and LL rings in the disk during accretion.

**Acknowledgment**

We thank T. Dunn, T. Burbine and F. DeMeo for their careful reviews.